\documentclass[11pt]{article}
\usepackage{amssymb,amsmath,cite,geometry,graphicx,url,mathrsfs}
\catcode`\@=11

\@addtoreset{equation}{section}
\def\theequation{\arabic{section}.\arabic{equation}}
     
\catcode`\@=11
\def\thesection{\arabic{section}}

\def\appendix{\setcounter{section}{0}
        \def\thesection{Appendix.}
        \def\theequation{\Alph{section}.\arabic{equation}}}
\def\section{\@startsection{section}{1}{\z@}{3.5ex plus 1ex minus
   .2ex}{2.3ex plus .2ex}{\large\bf}}
     
\long\def\@makefntext#1{\parindent 0cm\noindent
\hbox to 1em{\hss$^{\@thefnmark}$}#1}

\newcommand{\captionfonts}{\small}
\makeatletter  
\long\def\@makecaption#1#2{%
  \vskip\abovecaptionskip
  \sbox\@tempboxa{{\captionfonts #1: #2}}%
  \ifdim \wd\@tempboxa >\hsize
    {\captionfonts #1: #2\par}
  \else
    \hbox to\hsize{\hfil\box\@tempboxa\hfil}%
  \fi
  \vskip\belowcaptionskip}
\makeatother   

\begin{document}
\begin{titlepage}
\vspace{.5in}
\begin{flushright}
June 2021\\  
revised October 2021\\
 \end{flushright}
\vspace{.5in}
\begin{center}
{\Large\bf
 Midisuperspace foam\\[1ex]
  and the cosmological constant}\\  
\vspace{.4in}
{S.~C{\sc arlip}\footnote{\it email: carlip@physics.ucdavis.edu}\\
       {\small\it Department of Physics}\\
       {\small\it University of California}\\
       {\small\it Davis, CA 95616}\\{\small\it USA}}
\end{center}

\vspace{.5in}
\begin{center}
{\large\bf Abstract}
\end{center}
\begin{center}
\begin{minipage}{4.5in}
{\small
Wheeler's conjectured ``spacetime foam''---large quantum fluctuations of spacetime 
at the Planck scale---could have important implications for quantum gravity, perhaps
even explaining why the cosmological constant seems so small.  Here I explore this 
problem in a midisuperspace model consisting of metrics with local spherical symmetry.   
Classically,  an infinite class of ``foamy'' initial data can be constructed, in which
cancellations between expanding and contracting regions lead to a small average 
expansion even if $\Lambda$ is large.  Quantum mechanically, the model admits corresponding 
stationary states, for which the probability current is also nearly zero.  These states appear 
to describe a self-reproducing spacetime foam with very small average expansion, 
effectively hiding the cosmological constant.}
\end{minipage}
\end{center}
\end{titlepage}
\addtocounter{footnote}{-1}

\section{Spacetime foam and the cosmological constant}

More than 65 years ago, Wheeler suggested that quantum uncertainties in the metric should 
be of order one at the Planck scale, leading to large fluctuations in geometry and (perhaps) 
topology,  which he called ``spacetime foam'' \cite{Wheeler}.   While this idea has
continued to generate  interest, 
it has been notoriously hard to investigate  quantitatively.  My goal here is to develop a 
model simple enough to allow calculations, but still complex enough to describe a large
assortment of foam-like quantum states.

More specifically, one aim is to test a proposal  that spacetime foam might help solve
the cosmological constant problem.  In \cite{Carlip1}, I suggested that our Universe 
might in fact have a large cosmological constant $\Lambda$, whose effects are
``hidden''  in Planck scale fluctuations.   I showed that for a large class of initial data, 
the expansion and shear---the observational signature of a cosmological constant---average 
to zero over small regions even if $\Lambda$ is large.  A slightly oversimplified picture 
is that a cosmological constant can cause either expansion or contraction; using a 
construction of Chrusciel, Isenberg, and Pollack \cite{Chrusciel,Chruscielb}, one can sew 
together Planck-scale regions with random initial data in a way that naturally leads to 
large cancellations.  

A key question is whether such cancellations persist during evolution.  This is fundamentally 
a matter for quantum gravity, and a full treatment lies beyond our current capabilities.  
There is, however, a partial step: we can explore mini- and midisuperspace 
models,\footnote{Conventionally a minisuperspace keeps only finitely many degrees of 
freedom, while a midisuperspace has an infinite, although still restricted, set.}
in which many of the degrees of freedom are frozen out.   Such models have
been used to study a range of issues in quantum gravity, and while 
the results are rarely conclusive, they can be strongly suggestive.

Here I will focus on a midisuperspace consisting of geometries with local spherical symmetry.   
When combined with the ``dust time'' of Brown and Kucha{\v r} \cite{Brown,Marolf,Husain},
the Wheeler-DeWitt equation takes a Schr{\"o}dinger-like form, and can be solved by WKB
methods.  We shall see that there are stationary states in which a foamlike structure and  
small average expansion are preserved in time.  Moreover, while regularization ambiguities
exist,  ``foamy'' spacetimes seem to occur with high probabilities.  In this setting, 
at least, a cosmological constant may indeed be hidden in spacetime foam.

\section{Local spherical symmetry}

The setting for locally spherically symmetric midisuperspace has been studied 
extensively by Morrow-Jones, Witt, and Schleich \cite{Witt1,Witt2,Witt3}.  At first sight, this
symmetry requirement may seem too strong: it is widely believed that  spherical symmetry 
with $\Lambda>0$ leads inevitably to  Schwarzschild--de Sitter space.  But while Birkhoff's 
theorem implies that a spherically symmetric vacuum spacetime must be \emph{locally} isometric 
to some region of Schwarzschild-de Sitter space, local patches can be sewn together to form a 
spacetime that looks  drastically different  \cite{Witt3}.  In particular, we will be able to construct 
explicit initial data containing both expanding and contracting regions.

Here, for simplicity, I will assume a positive cosmological constant,
 and specialize to the spatial topology $S^1\times S^2$.  Recall that to construct a  
manifold with this topology, we start with a solid three-ball; cut out a ball from its center to form a 
manifold $[0,1]\times S^2$; and identify the boundaries $\{0\}\times S^2$ and $\{1\}\times S^2$.  
To build ``foamy'' spacetimes, 
we will take this construction a step further, splitting space into an onion-like 
sequence of concentric shells, each with its own geometry.

Following \cite{Witt1}, let us start with the initial value formalism, with
geometric data consisting of a spatial metric $q_{ij}$ and its conjugate momentum $\pi^i{}_j$.  
These cannot be chosen arbitrarily, but must solve the momentum and Hamiltonian constraints
\begin{subequations}
\begin{align}
&\mathscr{P}_j = {}^{\scriptscriptstyle(3)}\nabla_i\pi ^i{}_j = 0 \label{a0a}\\
&\mathscr{H} = \frac{2\kappa^2}{\sqrt{q}}\left(\pi^i{}_j\pi^j{}_i - \frac{1}{2}\pi^2\right) 
    - \frac{1}{2\kappa^2}\sqrt{q}\left({}^{\scriptscriptstyle(3)}\!R - 2\Lambda\right) = 0 .
 \label{a0b}
 \end{align}
 \end{subequations}
 (I use the conventions of \cite{Carlip3}; in particular, $\kappa^2=8\pi G$.)  On a hypersurface of 
 constant time, a general locally spherically symmetric metric may be written in the form
\begin{align}
ds^2 = h^2d\psi^2 + f^2(d\theta^2 + \sin^2\theta\, d\varphi^2)
\label{a1}
\end{align}
where $h$ and $f$ are functions of $\psi$ and $t$.  The spatial scalar curvature is 
\begin{align}
{}^{\scriptscriptstyle(3)}\!R 
   = \frac{1}{f^2}\left[ -\frac{4f}{h}\left(\frac{f'}{h}\right)^\prime -\frac{2f^{\prime2}}{h^2} + 2\right] ,
\label{a2}
\end{align}
where a prime denotes a derivative with respect to $\psi$.  Ref.\ \cite{Witt1} makes an added
coordinate choice $h(\psi,t)=a(t)$, $f(\psi,t)=a(t){\tilde f}(\psi)$; this simplifies the classical solutions,
but complicates the quantum treatment by gauge-fixing one of the diffeomorphism constraints.

The nonvanishing canonical momenta  are
\begin{align}
\pi^\theta{}_\theta = \pi^\varphi{}_\varphi = \sin\theta\,Q, \quad \pi^\psi{}_\psi = \sin\theta\,P ,
\label{a3}
\end{align}
where $P$ and $Q$ are functions of $\psi$ and $t$.  (The $\sin\theta$ factors appear because
the momenta are tensor densities).  Geometrically, $P$ and $Q$ are extrinsic curvatures, fractional
rates of expansion; explicitly,
\begin{align}
Q = \frac{1}{2\kappa^2}hf^2\left(K^\theta{}_\theta + K^\psi{}_\psi\right), \quad
P = \frac{1}{\kappa^2}hf^2K^\theta{}_\theta .
\label{a3a}
\end{align}
The only nontrivial momentum constraint is
 \begin{align}
 \mathscr{P} = P' - \frac{h'}{h}P -\frac{2f'}{f}Q = 0 \ \Rightarrow\ Q = \frac{f}{2f'}\left(P' - \frac{h'}{h}P\right) ,
 \label{a4}
 \end{align}
 while the Hamiltonian constraint becomes, after a little algebra,
 \begin{align}
 \mathscr{H} = \frac{h}{f'}\frac{d\ }{d\psi}\left[-\kappa^2\frac{P^2}{fh^2} 
   + \frac{1}{\kappa^2}\left(\frac{ff^{\prime2}}{h^2} - f\right) + \frac{\Lambda}{3\kappa^2}f^3\right]  = 0.
\label{a5}
\end{align}
This simple total derivative structure was first noted in \cite{Witt1}.

Given a metric $(h,f)$, we can now solve (\ref{a5}) for $P$,
\begin{align} 
P = \pm \frac{1}{\kappa^2}\left[f^2h^2\left(\frac{f^{\prime2}}{h^2}-1\right) 
   + \frac{\Lambda}{3}f^4h^2 + \gamma fh^2\right]^{1/2} 
\label{a6}
\end{align}
where $\gamma$ is an integration constant.  As also noted in \cite{Witt1}, this constant is a mass:
if we look at a slice $P=0$ and gauge fix to $f=\psi$ (allowed locally, though not globally), we see that
\begin{align}
h^2 = \left(1-\frac{\gamma}{\psi} - \frac{\Lambda}{3}\psi^2\right)^{-1},
\label{a7}
\end{align}
a piece of the usual Schwarzschild-de Sitter metric with $\gamma=2Gm$.

Note that the value of the integration constant $\gamma$ can have a drastic effect on the spacetime
geometry \cite{Lake,SW}.  Here I will assume that
\begin{align}
0< \gamma < \frac{2}{3\sqrt{\Lambda}}  .
\label {a7x}
\end{align}
For the complete Schwarzschild-de Sitter metric, this is the condition for the existence of two horizons,
that is, for a black hole that is smaller than the cosmological horizon.  As we will see shortly, it is also
necessary for the existence of a Cauchy surface with vanishing extrinsic curvature, a requirement for 
attaching expanding and contracting regions to realize the ``spacetime foam'' of \cite{Carlip1}.

By standard existence theorems, the initial data 
$(h,f,P,Q)$ can now be evolved to form a maximal globally hyperbolic spacetime.  As expected,
the system has a time reversal symmetry: if  the data $(h,f,P,Q)$ are admissible, so are
$(h,f,-P,-Q)$.  Since $P$ and $Q$ determine the expansion and shear, for any expanding solution 
there is a corresponding contracting one.

Of course, there is no guarantee that the resulting spacetimes are geodesically complete.  On
the contrary, regions of the initial surface can be isometric to regions of the interior of a 
Schwarzschild-de Sitter black hole, so one might generically expect singularities to form, as
happens in the more general setting \cite{Burkhartb}.  Evolution thus requires a quantum 
treatment.

\section{Sewing \label{sew}}

Ref.\ \cite{Carlip1} focused on manifolds formed by ``sewing'' elementary pieces
to build a model of spacetime foam.   An almost identical procedure exists in the spherically
symmetric setting \cite{Witt1}.  Choose coordinates in which $h$ is constant 
in a small region around $\psi=\psi_0$.  Pick two manifolds $M_1$ and $M_2$, 
with arbitrary functions $f_1(\psi)$ and $f_2(\psi)$.  Let $u_\epsilon$ be a smoothed step
function, interpolating between $u_\epsilon(\psi)=0$ when $\psi<-\epsilon$ and 
$u_\epsilon(\psi)=1$ when $\psi>\epsilon$.  

Now define
\begin{align}
f(\psi) = [1-u_\epsilon(\psi-\psi_0)]f_1(\psi) + u_\epsilon(\psi-\psi_0)f_2(\psi)  .
\label{b1}
\end{align}
This ``sewn''  $f$ looks like $f_1$ for $\psi\in(-\infty,\psi_0-\epsilon)$ and like $f_2$ for
$\psi\in(\psi_0+\epsilon,\infty)$, with a smooth interpolation between.  Since the momenta $P$
and $Q$ are determined locally from $f$ and $h$, the full initial data will sew together in the same way.
Topologically, this process is a connected sum $M_1\#M_2$; geometrically, it is very close to the  
general construction of \cite{Chrusciel,Chruscielb}.

This isn't yet quite good enough.  The functions $f_1$ and $f_2$ each 
determine two sets of data, $(h,f_1,\pm P_1,\pm Q_1)$ and $(h,f_2,\pm P_2,\pm Q_2)$.  
We would like to sew any combination, for instance attaching an expanding region to a contracting 
region.  But for $P$ and $Q$ to change sign, they must go through zero, a condition that 
is not automatic.

To allow such a sign change, we construct an extra ``neck'' $N$, an intermediate manifold 
in which $P$ and $Q$ go to zero on a central two-sphere.  We can then sew $M_1$ and 
$M_2$ via this neck, $M_1\#N\#M_2$, with either choice of momenta on each side.
We specify $N$, say with $\psi\in(-\delta,\delta)$, by demanding that
 \begin{align} 
 {f}'(0) =0, \  f^{\prime\prime}(0) >0 ,\  P(0) = Q(0) = 0 .
\label{b2}
\end{align}

 The condition $P(0)=0$ is easy: by (\ref{a6}), it simply requires that
\begin{align}
\frac{\Lambda}{3}f_0{}^3 = f_0-\gamma ,
\label{b3}
\end{align}
where $f_0 = f(0)$.   Note, though, that (\ref{b3}) has real, positive roots only if the inequality
(\ref{a7x}) is satisfied \cite{Lake}.  If the inequality is violated, $P$ can never go through zero,
and it becomes impossible to join expanding and contracting regions, at least within the constraint
of local spherical symmetry.

Setting $Q(0)=0$ is harder: eqn.\ (\ref{a4}) for $Q$ has a factor $f'$ in the denominator, which 
goes to zero at the center of the neck. To cancel this term, we must demand that
$P(\psi)$ be of order $\psi^3$ near $\psi=0$.  A  straightforward calculation yields 
the requirement
\begin{align}
f(\psi) 
&= f_0 + \left( 1 - \Lambda f_0^2\right)f_0%
     \left(\frac{h^2\psi^2}{4f_0{}^2}\right)\left(1 - \frac{h^2\psi^2}{12f_0{}^2}\right)
    + \mathcal{O}\left(\psi^6\right) 
    \quad \hbox{with}\  1-\Lambda f_0^2 > 0  .
\label{b4}
\end{align}
This restriction is not quite as stringent as it
may seem: the $\mathcal{O}\left(1\right) $ and $\mathcal{O}\left(\psi^2\right)$ terms are
already fixed by the demand that $Q$ remain finite at $f'=0$.

By the constraint (\ref{a0b}), the scalar curvature 
${}^{\scriptscriptstyle(3)}\!R$ is large at $\psi=0$, since the extrinsic curvature vanishes there.  But 
the curvature falls off quickly away from the center of the neck:
\begin{multline}
{}^{\scriptscriptstyle(3)}\!R = 2\Lambda - \frac{8\kappa^4}{h^2f^4}\left(PQ-\frac{1}{4}P^2\right)
   = 2\Lambda - \frac{2\kappa^4}{3h^2}\frac{1}{f_0^3f^{\prime\prime}(0)}P_3{}^2\psi^4 
   + \mathcal{O}\left(\psi^5\right)\\[.4ex]
   \hbox{where}\ P = \frac{1}{6}P_3\psi^3 + \mathcal{O}\left(\psi^4\right) .
\label{b4a}
\end{multline}
 
We can now construct the promised ``onion'' initial data, with a mix of expanding and 
contracting layers.  Choose a $K$-layer sequence $\psi_0<\psi_1<\dots<\psi_K$, where $\psi_K$ will be 
identified with $\psi_0$ to create a closed $S^1\times S^2$.  In each layer $\psi\in(\psi_n,\psi_{n+1})$, 
choose functions $(h_n,f_n)$, and randomly pick the sign of the corresponding $\pm(P_n,Q_n)$.  
Sew successive layers along necks as described above.  For a configuration with enough layers,
the random choice of signs means that $P$ and $Q$ will typically average to near zero, even if 
$\Lambda$ is large.  Thus, as in \cite{Carlip1}, we have hidden the initial cosmological constant, but 
now in a setting in which we can address the quantum theory much more concretely.

\section{The problem of time \label{time}}

To investigate the quantum evolution of this system, we must first confront a basic
puzzle, the notorious ``problem of time'' \cite{Kuchar,Isham}.   Quantum gravity is a relational theory, 
and time evolution can only be described relative to other degrees of freedom.  To describe
evolution, we must first choose a ``clock.''

In some expanding minisuperspace models \cite{Misner}, and perhaps more generally \cite{expand}, one 
can use spatial volume as a clock.  Here, though, our spacetimes have some regions that expand 
while others contract.  One can often use ``York time'' \cite{York}, in which the trace of 
the extrinsic curvature---the local Hubble constant---serves as a clock.  But here, again, 
we are interested in cases in which the extrinsic curvature doesn't even have a fixed sign at constant 
time.  A more general choice of time, the mean curvature flow \cite{MCF}, fails for the same reason.  
In some cosmological spacetimes, ``time since the big bang'' can serve as a clock \cite{costime}, 
but it is not obvious that the spacetimes considered here have a suitable initial big bang singularity.

An alternative,  developed by Brown and Kucha{\v r} \cite{Brown,Marolf}, uses
congruences of test particles, ``dust,'' as a reference system.\footnote{Footnote 1 of \cite{Torre} gives 
a history of this idea, which dates back to Einstein's ``mollusc of reference.''}   This is not ideal---it 
would be preferable to describe the evolution purely in terms of a gravitational variable---but in the 
present setting I don't know how to define such a ``clock.''
Here I will use a slightly simpler version of dust time, due to Husain and Paw{\l}owski \cite{Husain} 
(see also \cite{Torre}), in which a space-filling congruence of irrotational dust is used to specify 
time evolution.  The ``cloud of clocks'' is introduced with an action
\begin{align}
I_{\hbox{\tiny dust}} = \frac{1}{2}\int d^4x\sqrt{-g}\,\rho\left(g^{ab}\partial_aT\partial_bT -1\right) ,
\label{c1}
\end{align}
where $\rho$ is a Lagrange multiplier.  It is easy to see that the variation of $I_{\hbox{\tiny dust}}$
yields the usual stress-energy tensor for noninteracting irrotational matter.  Carrying out a
standard canonical decomposition of $I_{\hbox{\tiny grav}}+I_{\hbox{\tiny dust}}$, one finds  
constraints
\begin{subequations}
\begin{align}
&\mathscr{P}_i = \mathscr{P}_i^{\hbox{\tiny grav}} + \mathscr{P}_i^{\hbox{\tiny dust}} = 0  \label{c2a}\\
&\mathscr{H} = \mathscr{H}^{\hbox{\tiny grav}} + \mathscr{H}^{\hbox{\tiny dust}} = 0  ,
\label{c2b}
\end{align}
\end{subequations}
where $\mathscr{H}^{\hbox{\tiny grav}}$ and $\mathscr{P}_i^{\hbox{\tiny grav}}$ are given by 
(\ref{a0a})--(\ref{a0b}) and
\begin{subequations}
\begin{align}
&\mathscr{P}_i^{\hbox{\tiny dust}} = -p_T\partial_iT  \label{c3a}  ,\\
&\mathscr{H}^{\hbox{\tiny dust}} 
  = \left(p_T{}^2 +  q^{ij}\mathscr{P}_i^{\hbox{\tiny dust}}\mathscr{P}_j^{\hbox{\tiny dust}}\right)^{1/2} ,
\label{c3b}
\end{align}
\end{subequations}
where $p_T$ is the momentum conjugate to $T$.

As shown in \cite{Brown}, one can use these constraints in a ``many-fingered time'' quantization, 
in which the wave function evolves as a functional of $T$.  But one can also gauge fix time by choosing
$T=t$.  Physically, this amounts to using proper time along the dust worldlines as a time coordinate.
With this choice, $\mathscr{P}_i^{\hbox{\tiny dust}} = 0$, and the constraints reduce to
\begin{subequations}
\begin{align}
&\mathscr{P}_i^{\hbox{\tiny grav}} = 0 \label{c4a}\\
&p_T = \mathscr{H}^{\hbox{\tiny grav}} \label{c4b}    .
\end{align}
\end{subequations}

Eqn.\ (\ref{c4b}) has a slightly peculiar structure.  The left-hand side is independent of position, since 
we have chosen a gauge in which $T$ depends only on time, but the right-hand side is certainly  
position-dependent.  This is actually sensible: the zero mode of $\mathscr{H}^{\hbox{\tiny grav}}$ is now a
true Hamiltonian, while the remaining position-dependent modes continue to act as constraints, much 
like the situation in unimodular gravity \cite{unimodular}.

\section{Wheeler-DeWitt quantization \label{WQ}}

To quantize this system a la Wheeler and DeWitt \cite{WdW}, we rewrite the momentum
and  Hamiltonian constraints as operators, with canonical momenta replaced by 
(functional) derivatives.  To find the appropriate conjugates, we start with the symplectic current
\begin{align}
\omega = \int\!d^3x\,\pi^{ij}\delta q_{ij} = 8\pi\int\!d\psi\,\left(\frac{2}{f}Q\,\delta f + \frac{1}{h}P\,\delta h\right) ,
\label{d0}
\end{align}
from which can we read off
\begin{subequations}
\begin{align}  
P = \frac{i}{8\pi}h\frac{\delta\ }{\delta h}, \quad  Q =  \frac{i}{16\pi}f\frac{\delta\ }{\delta f} . 
\label{d1a}
\end{align}
\end{subequations}
The spatial volume $\hat V$ and the mean curvature $\hat K$ (the trace of the extrinsic curvature) become
\begin{subequations}
\begin{align}  
&{\hat V} = \int\!d^3x\,\sqrt{q} = 4\pi\int\!d\psi\,hf^2  \label{d2a}\\[.5ex]
&{\hat K} = -\frac{\kappa^2}{\sqrt{q}}\,{\hat\pi}
     = -\frac{i\kappa^2}{8\pi}\frac{1}{hf^2}\left( h\frac{\delta\ }{\delta h} + f\frac{\delta\ }{\delta f}\right) ,\label{d2b}
\end{align}
\end{subequations}
and satisfy $\displaystyle [{\hat K},{\hat V}] = -{3i\kappa^2}/{2}$ .
 
The momentum constraint (\ref{a4}) now becomes
\begin{align}
\mathscr{\hat P} = \frac{i}{8\pi}\left(h\partial_\psi\frac{\delta\ }{\delta h} - f'\frac{\delta\ }{\delta f}\right) .
\label{d2}
\end{align}
Acting on $h$ and $f$,  $\mathscr{\hat P} $ generates spatial diffeomorphisms, and invariant
wave functions can be built from integrals of the form
\begin{align}
F[h,f] = \int\!d\psi\,h L[f,Df,D^2f,\dots] \qquad\hbox{with $\displaystyle D = \frac{1}{h}\frac{d\,}{d\psi}$}.
\label{d3}
\end{align}
The Hamiltonian constraint (\ref{a5}) is a bit more complicated, but can be reduced to the form
\begin{align}
\mathscr{\hat H} = \frac{h}{f'}\frac{d\ }{d\psi}\left[
    \frac{\kappa^2}{64\pi^2}\frac{1}{f}\frac{\delta^2\ }{\delta h^2}  
    + \frac{1}{\kappa^2}f\left(\frac{f^{\prime 2}}{h^2} -1\right) + \frac{\Lambda}{3\kappa^2}f^3\right] ,
\label{d4}
\end{align}
while the momentum $p_T$ of (\ref{c4b}) becomes
\begin{align}
p_T = i hf^2\frac{d\ }{dT} 
\label{d5}
\end{align}
($hf^2$ is the volume measure $\sqrt{q}$).  The Wheeler-DeWitt equation---the operator version of
(\ref{c4b}), acting on a wave function $\Psi$---is thus
\begin{align}
if^2f'\frac{d\Psi}{dT} = \frac{d\ }{d\psi}\left[
    \frac{\kappa^2}{64\pi^2}\frac{1}{f}\frac{\delta^2\ }{\delta h^2}  
    + \frac{1}{\kappa^2}f\left(\frac{f^{\prime 2}}{h^2} -1\right) + \frac{\Lambda}{3\kappa^2}f^3\right] \Psi .
\label{d6}
\end{align}

As in the classical case, (\ref{d6}) can be integrated.   The derivative $d\Psi/dT$ is 
independent of position---$\Psi$ is a functional of $h$ and $f$, not a function of $\psi$---so
the left-hand side of (\ref{d6}) is a total derivative.  Hence  
\begin{align}
i\frac{d\Psi}{dT} = \left[ \frac{3\kappa^2}{64\pi^2}\left(\frac{1}{f^2}\frac{\delta\ }{\delta h}\right)^2 
    + \frac{3}{\kappa^2}\frac{1}{f^2}\left(\frac{f^{\prime 2}}{h^2} -1\right) + \frac{\Lambda}{\kappa^2}
    + \frac{3\gamma}{\kappa^2 f^3}\right]\Psi
    = {\hat H}\Psi ,
\label{d7}
\end{align}
where $\gamma$ is again an integration constant.  This is the fundamental equation we must solve
to understand quantum midisuperspace.

\subsection{The wave function}

Eqn.\ (\ref{d7}) is a Schr{\"o}dinger-type equation, and there is no difficulty in principle
in interpreting $|\Psi|^2$ as a probability density on midisuperspace \cite{Kuchar}.  In particular, 
the time derivative
\begin{align}
 \frac{d(\Psi^*\Psi)}{dT} = - \frac{3i\kappa^2}{64\pi^2}\frac{\delta\ }{\delta h}%
    \left[\frac{1}{f^4}\left(\Psi^*\frac{\delta\Psi}{\delta h} - \Psi\frac{\delta\Psi^*}{\delta h}\right)\right] 
\label{da1}
\end{align}
is a total (functional) derivative, so formally, the norm
$$\int [df][dh]\Psi^*\Psi$$
is conserved, at least for suitable boundary conditions for midisuperspace.  

There is a subtlety, though.  The wave function $\Psi$ should really be defined on ``reduced midi\-superspace,'' 
the space of symmetric metrics modulo diffeomorphisms.  To   properly define a norm on this quotient 
space, one must gauge fix the inner product to avoid overcounting diffeomorphism-equivalent 
configurations \cite{Woodard}.  This is a difficult task, since one must account for not only the 
spatial diffeomorphisms, but also the symmetries generated by the non-zero modes of the Hamiltonian 
constraint.  Partial results are given in the Appendix; for now, I will avoid this issue, and merely use 
$|\Psi|^2$  to determine relative probabilities, with the understanding that the inner product measure 
may give corrections.

It is worth reiterating that $\Psi$ is a functional of the metric, not a function of space.  Its role is to 
determine probabilities of configurations $(h,f)$.  Of course, the probability of any given 
metric occurring will depend on its spatial form, and we shall see that a wave function can imply ``a 
high probability for metrics with spatial characteristic X.''   But this will only make sense if the ``spatial 
characteristic X'' is expressed in a diffeomorphism-invariant way, one that does not refer to any
particular values of the coordinates.  
 
\subsection{The probability current}

As a Schr{\"o}dinger-type equation, (\ref{d7}) should also admit a probability current.
Recall that in the WKB approximation in ordinary quantum 
mechanics, the probability current distinguishes genuinely time-independent configurations
(e.g., bound states) from steady state descriptions of a secretly time-dependent configurations
(e.g., plane waves scattering off a potential barrier).  We shall see that the same is true here.

It is easy to see that the current
\begin{align}
J[f,h;x] = \frac{3i\kappa^2}{64\pi^2}%
    \left(\Psi^*\frac{1}{f^2}\frac{\delta\Psi}{\delta h} - \Psi\frac{1}{f^2}\frac{\delta\Psi^*}{\delta h}\right) .
\label{d8}
\end{align}
obeys a sort of continuity equation on midisuperspace,
\begin{align}
 \frac{d(\Psi^*\Psi)}{dT} + \frac{1}{f^2}\frac{\delta J}{\delta h} = 0 .
\label{d9}
\end{align}
But $J$ is not an observable: it is not annihilated by the momentum constraint $\mathscr{\hat P}$.
We can project $J$ onto reduced midisuperspace by taking a  volume average, or equivalently 
forming a group average over the spatial diffeomorphisms \cite{Marolfb},
\begin{align}
\langle J\rangle = \frac{1}{V}\int\!d^3x\,hf^2J ,
\label{d10}
\end{align}
but this average no longer obeys an obvious continuity equation.  

To do better, consider the functions
\begin{align}
\varphi_n(\psi) = \exp\left\{\frac{8\pi^2 i n}{V}\int^\psi\!\! f^2h\,d\psi'\right\}  .
\label{d11}
\end{align}
These are orthogonal and complete:
\begin{align}
& \int hf^2\varphi_m^*\varphi_n\,d\psi = \frac{V}{4\pi}\delta_{mn} , \nonumber\\[1ex]
&\sum_n \varphi_n^*(\psi')\varphi_n(\psi) = \frac{V}{4\pi hf^2}\,\delta(\psi-\psi')  .
\label{d12}
\end{align}
If we now project 
\begin{align}
D_n = \frac{1}{V}\int\!d^3x\,hf^2\varphi^*_n\left(\frac{1}{f^2}\frac{\delta\ }{\delta h}\right) , \ \ 
J_n = \int\!d^3x\,hf^2\varphi_n\,J ,
\label{d13}
 \end{align}
 a bit of work shows that
 \begin{align}
 \sum_n D_nJ_n = \frac{1}{V}\int\!d^3x\,hf^2\left(\frac{1}{f^2}\frac{\delta J}{\delta h}\right) = -  \frac{d(\Psi^*\Psi)}{dT} ,
 \label{d14}
 \end{align}
where the last equality comes from (\ref{d9}).
$D_n$ and $J_n$ are spatial invariants, that is, that they are annihilated by the
momentum constraint.\footnote{$D_n$ and $J_n$ are not annihilated by the Hamiltonian constraint; a further 
projection may be needed to construct spacetime invariants, but that lies beyond the scope of this paper.}  
The $D_n$ may be viewed as the components of the gradient in our (infinite-dimensional) midisuperspace, 
making (\ref{d14}) a standard continuity equation.   
 
\section{Stationary states and the WKB approximation}

We next look for stationary states
\begin{align}
\Psi[f,h;T] = {\tilde\Psi}[f,h]e^{-iET}  .
\label{e1}
\end{align}
For states of this form, the Wheeler-DeWitt equation (\ref{d7}) becomes
\begin{align}
\left[ \frac{\kappa^2}{64\pi^2}\left(\frac{1}{f^2}\frac{\delta\ }{\delta h}\right)^2 
    + \frac{1}{\kappa^2}\frac{1}{f^2}\left(\frac{f^{\prime 2}}{h^2} -1\right) + \frac{\tilde\Lambda}{3\kappa^2} 
    + \frac{\gamma}{\kappa^2 f^3}\right]{\tilde\Psi} = 0  ,
\label{e2}
\end{align}
where 
\begin{align}
{\tilde\Lambda} = \Lambda - \kappa^2 E .
\label{e2a}
\end{align}  
This equation is identical to the original gravitational Hamiltonian constraint, except that the
cosmological constant is shifted by the energy, again reminiscent of unimodular gravity
\cite{unimodular}.  This shift in $\Lambda$ is physical---it is a backreaction of our ``cloud of clocks'' on 
the spacetime---and to see purely gravitational properties, we should limit ourselves to states with relatively 
small energies.   For a Planck-scale $\Lambda$, though, this is a rather mild restriction.

To better understand these states, let us consider a WKB approximation,
\begin{align}
{\tilde\Psi} = A e^{iS}  .
\label{e3}
\end{align}
The first order WKB equation,
\begin{align}
\frac{\kappa^2}{64\pi^2}\left(\frac{\delta S}{\delta h}\right)^2 
    = \frac{1}{\kappa^2} f^2\left(\frac{f^{\prime 2}}{h^2} -1\right) + \frac{\tilde\Lambda}{3\kappa^2} f^4
    + \frac{\gamma}{\kappa^2} f ,
\label{e4}
\end{align}
can be solved exactly:
\begin{align}
S = \frac{8\pi}{\kappa^2}\int\!d\psi\,\sigma[h,f;\psi]ff'\left\{\sqrt{1+\beta h^2} - \tanh^{-1}\sqrt{1+\beta h^2} \right\} ,
\label{e5}
\end{align}
where
\begin{align}
\beta = \frac{f^2}{f^{\prime2}}\left(\frac{\tilde\Lambda}{3} - \frac{1}{f^2} + \frac{\gamma}{f^3}\right)
\label{e6}
\end{align}
and $\sigma$ is a functional of $h$ and $f$ and a function of $\psi$ such that
\begin{align}
\left\{ \begin{array}{ll}\sigma^2 = 1 \qquad&\hbox{almost everywhere}\\[.5ex]
  \partial_\psi\sigma =0 &\hbox{unless}\  1+\beta h^2 = 0  .
   \end{array}\right.
\label{e7}
\end{align}

The factor $\sigma$ requires a bit of explanation.   As always in the WKB approximation, the 
phase $S$ is determined only up to sign. But  eqn.\ (\ref{e4}) holds pointwise, and thus admits solutions 
in which this  sign, $\sigma$, can vary with $f$, $h$, and $\psi$.  The choice is not completely 
free, though: we must still require that $S$ be annihilated by the momentum constraint.  If the position 
dependence of $\sigma$  only arose implicitly from its dependence on $h$ and $f$, then $S$ 
would be of the form (\ref{d3}),  and we would automatically have $\mathscr{\hat P}S = 0$.  But any 
explicit dependence of $\sigma$ on $\psi$ introduces a new term in $\mathscr{\hat P}S$,   proportional to 
$(1+\beta h^2)^{1/2}\,\partial_\psi\sigma$.  The constraints require this term to vanish, leading to the 
second condition in (\ref{e7}).

To understand the physical significance of this condition, we can rewrite (\ref{e6}) as
\begin{align}
1+\beta h^2 = \frac{\kappa^4{\tilde P}^2}{f^2f^{\prime2}} ,
\label{e9}
\end{align}
where 
\begin{align}
{\tilde P} = \frac{1}{\kappa^2}\left[f^2h^2\left(\frac{f^{\prime2}}{h^2}-1\right) 
   + \frac{{\tilde\Lambda}}{3}f^4h^2 + \gamma fh^2\right]^{1/2}   .
\label{e9a}
\end{align}
Comparing to (\ref{a6}), we see that ${\tilde P}$ is essentially the classical momentum 
$\pi^\psi{}_\psi$, though with a shifted cosmological constant.  Hence the sign 
of $S$ can change only when the ${\tilde P}$ goes through zero, the quantum version of the the 
classical sewing condition of section \ref{sew}.  There is also an analog of the inequality (\ref{a7x})
for $\gamma$.  For large $f$, $\beta$ is positive, while for $\tilde P$ to go through zero, $\beta$ 
must be negative.  But  (\ref{e6}) involves the same cubic as (\ref{b3}), and $\beta$ 
can change sign only if the inequality (\ref{a7x}) is satisfied, albeit again with a 
shifted $\Lambda$.

This means, in particular, that for suitable $\gamma$, the quantum theory allows such sewing.   More 
precisely, let $(h,f)$ be a metric configuration that, as a classical metric, describes layers joined by 
necks in which the extrinsic curvature goes through zero.  A choice of $\sigma$ for this configuration is 
then equivalent to a choice of the sign of the extrinsic curvature in each layer.  Just as this sign 
is not determined classically by the constraints, it is not determined quantum mechanically by the
 Wheeler-DeWitt equation: different choices of $\sigma$ give different wave functions $\Psi_\sigma$. 

We can now explore our WKB wave functions in various regions of midisuperspace.  First, consider 
the case of large $f$---that is, specialize to metrics in which $f$ is large in some region of space, 
and look at the contribution of that spatial region to the integral (\ref{e5}).  In such a region,
${\tilde P}\gg0$, so the sign $\sigma$ is fixed, and $\beta$ is dominated by the cosmological 
constant, yielding  
\begin{align}
S \sim \frac{2}{\kappa^2}\,\sigma\biggl(\frac{\tilde\Lambda}{3}\biggr)^{1/2}V  .
\label{e8}
\end{align}
The mean curvature operator (\ref{d2b})---the local Hubble constant---then has a simple action,
\begin{align}
{\hat K}e^{iS} \approx 3\sigma \biggl(\frac{\tilde\Lambda}{3}\biggr)^{1/2}\,e^{iS}  .
\label{e8a}
\end{align}
This is ordinary de Sitter behavior, and, as expected, the sign $\sigma$ determines whether a spatial
region is expanding or contracting.  The probability current (\ref{d13}) provides similar information:
\begin{align}
J_n = -\frac{3V}{4\pi}\sigma\biggl(\frac{\tilde\Lambda}{3}\biggr)^{1/2}\delta_{n0}  ,
\label{e8b}
\end{align}
so $\sigma$ determines the direction of flow of probability.  This is not unlike the WKB approximation 
in ordinary quantum mechanics, where a plane wave $e^{ikx}$ is formally a stationary state, but the 
probability current reveals a hidden dynamics.

But the wave function does not have its support only on such de Sitter-like regions.  Consider the
contribution of a multilayered ``foamy'' region.  From (\ref{d2b}) and (\ref{d2}),  the mean curvature acts 
on diffeomorphism-invariant states as
\begin{align}
{\hat K} = \frac{i\kappa^2}{8\pi}\frac{1}{f^2f'}\partial_\psi\left(f\frac{\delta\ }{\delta h}\right)  ,
\label{e11}
\end{align}
yielding 
\begin{align}
{\hat K}e^{iS} = \frac{1}{f^2f'}\partial_\psi\left(\frac{f^2f'}{h}\sigma\sqrt{1+\beta h^2}\right)e^{iS} 
  =   \frac{\kappa^2}{f^2f'}\partial_\psi\left(\frac{f}{h}\sigma{\tilde P}\right)e^{iS}   .
\label{e12}
\end{align}
As expected, the WKB wave function is not an eigenfunction of ${\hat K}$.  But  in for foamy
regions of midisuperspace, it is almost an eigenfunction of the spatially averaged mean 
curvature.  Indeed,
\begin{align}
\left(\frac{1}{V}\int\!d^3x\,\sqrt{q}{\hat K}\right) e^{iS} 
   = -\frac{4\pi\kappa^2}{V}\int\!d\psi\, \frac{f}{h}\left(\frac{h}{f'}\right)^\prime\sigma{\tilde P}\, e^{iS} .
\label{e13}
\end{align}
For a typical wave function, evaluated at a typical multilayered metric, the contribution of each layer will
come with a random sign $\sigma$.  Hence for a configuration with many layers, we expect extensive 
cancellation---just as in the classical case, the averaged mean curvature will be much smaller than it 
would be for a single de Sitter region.  

The evaluation of the probability current leads to an identical conclusion, perhaps even more clearly.  
The current is
\begin{align}
J_n = -\frac{3\kappa^2}{4}\int\!d^3x\,\sigma {\tilde P}\varphi_n ,
\label{e10}
\end{align}
and for foamy metrics the right-hand side will again average to a very small number.   Our
wave functions are thus stationary not only in the sense that they are independent of $T$, but
also in the sense that there is very little flow of probability within midisuperspace, so initial foamy
structures will tend to be preserved.
 
Of course, this leaves open the question of how common such multilayered foamy configurations
are.  An answer requires the next order WKB approximation, which we will turn to
shortly.  Meanwhile, there are a few other features of (\ref{e5}) that  deserve future exploration:
\begin{enumerate}
\item The integrand in (\ref{e5}) diverges when $\beta h^2=0$.  Comparing (\ref{e6}) to (\ref{a7}), we   
might suspect this to occur at some sort of horizon.  This is correct: the 
vanishing of $\beta h^2$ marks the location of a marginally trapped sphere, a midisuperspace 
version of the trapped surfaces that appear in the more general connected sum analysis \cite{Burkhartb}.
\item The integrand vanishes at $1+\beta h^2=0$, the classical neck.  But it also has zeroes at 
$1+\beta h^2 = -z_n^2$, where $z_n=\tan z_n$.  These are in a classically forbidden region,
but in principle they are additional sites at which expanding and contracting regions can join.
\item When $1+\beta h^2>0$, $\tanh^{-1}\sqrt{1+\beta h^2}$ has a constant imaginary part.  In a 
region between two successive zeroes of $1+\beta h^2$, though, $\sigma$ is constant, and the
imaginary part of $S$ is
$$\frac{4\pi^2 i}{\kappa^2}\int_{\psi_1}^{\psi_2}\!d\psi\,\sigma ff' = \frac{2\pi^2 i}{\kappa^2}\sigma[f^2(\psi_2)-f^2(\psi_1)]$$
with $1+\beta h^2 = 0$ at $\psi_1$ and $\psi_2$.  We saw earlier that at a ``classical'' zero of $1+\beta h^2$,
the value of $f$ is fixed by (\ref{b3}), so the contributions of the two endpoints cancel.  It is less clear
what happens at the nonclassical zeroes $1+\beta h^2 = -z_n^2$.
\end{enumerate}

\section{Next order WKB}

To say more about probabilities for ``foamy'' spacetimes, we will need to go to the next order
in the WKB approximation (\ref{e3}),
\begin{align}
2\frac{\delta S}{\delta h}\frac{\delta A}{\delta h} + A\frac{\delta^2S}{\delta h^2} = 0  .
\label{f1}
\end{align}
This equation involves two functional derivatives at a single point, leading to a well known
divergence.  For our midisuperspace model, we can write
\begin{align}
S = \int\!d\psi\,L ,
\label{f2}
\end{align}
where $L$ depends on $h$ but not $h'$.  Then
\begin{align}
\frac{\delta S}{\delta h(\psi)} = \frac{\partial L}{\partial h}(\psi), \ \
\frac{\delta^2 S}{\delta h(\psi)\delta h(\psi')} = \frac{\partial^2 L}{\partial h^2}(\psi)\delta(\psi-\psi') ,
\label{f3}
\end{align}
giving a factor $\delta(0)$ at $\psi=\psi'$.

There are several proposals for regulating this infinite factor.  In the original formulation of the 
Wheeler-DeWitt equation \cite{WdW}, DeWitt suggested that such $\delta(0)$ factors should 
be set to zero.  Here, that choice would make the first order WKB approximation exact.  
A ``volume average regularization'' \cite{Feng}, applied to this simple case, replaces 
$\delta(\psi-\psi')$ by 
$$\frac{h}{\ell}\int d\psi'\delta(\psi-\psi')$$
where $\ell = \int\!d\psi\, h$ is the length of the spatial $S^1$ factor.  
Heat kernel regularization \cite{Horiguchi} replaces the factor $\delta(\psi-\psi')$  
by a heat kernel $K(\psi,\psi';s)$.  In the $s\rightarrow0$ limit,  $K(\psi,\psi';s)$ becomes a delta function,
but if one first takes $\psi\rightarrow\psi'$, the divergences appear as a set of terms involving 
inverse powers of $s^{1/2}$, which can be individually regularized.  In the present context, the 
outcome is almost the same as the volume average: $\delta(\psi-\psi')$ becomes
$$\alpha h \int d\psi'\delta(\psi-\psi')$$
where $\alpha$ is an undetermined constant with dimensions of inverse length.
 
\subsection{Heat kernel regularization \label{heat}} 

Let us first consider heat kernel regularization.  The second  order WKB equation becomes
\begin{align}
\frac{1}{A}\frac{\delta A}{\delta h} 
   = -\frac{\alpha h}{2}\left(\frac{\partial L}{\partial h}\right)^{-1} \frac{\partial^2L}{\partial h^2} ,
\label{f4}
\end{align}
or with $S$ as in (\ref{e5}),  
\begin{align}
\frac{1}{A}\frac{\delta A}{\delta h} = \frac{\alpha}{2}\,\frac{1}{1+\beta h^2} .
\label{f5}
\end{align}
This can be solved in closed form:
\begin{align}
A=\exp\left\{ \frac{\alpha}{2}\int\!d\psi\,\frac{1}{\sqrt{\beta}}\tan^{-1}(\sqrt{\beta}h)\right\}  .
\label{f6}
\end{align}

We can now look at relative probabilities for behaviors of the metric---that is,  at features 
of the functions $(h,f)$ that lead to comparatively small or large vales of the amplitude $A$.

\begin{itemize}
\item
When $\beta h^2$ is large, the integrand in (\ref{f6}) is small, since $\tan^{-1}(\sqrt{\beta}h)$ 
is bounded by $\pi/2$.  By (\ref{e9}), this implies that metrics with large regions of high extrinsic 
curvature are suppressed.  If we abbreviate
\begin{align}
\left|\frac{\kappa^2{\tilde P}}{ff'}\right| =  \sqrt{1+\beta h^2} = \Delta
\label{f7x}
\end{align}
then for a region of length $\ell_1$,
\begin{align}
\ln A\bigl|_{\Delta\gg1} \ 
   \sim\ \frac{\alpha\pi}{4}\left\langle\Delta^{-1}\right\rangle\ell_1 \ll \frac{\alpha\ell_1}{2} ,
\label{f7a}
\end{align}
where the angle brackets denote a spatial average.
\item 
When $\beta h^2$ is small but positive, it is still the case that $\tan^{-1}(\sqrt{\beta}h) < \sqrt{\beta}h$, 
so
\begin{align}
\ln A\bigl|_{\Delta\sim1}\  \lesssim \ \frac{\alpha\ell_1}{2}  .
\label{f7b}
\end{align}
\item
When $\beta h^2<0$,  
\begin{align}
\frac{\alpha}{2}\int\!d\psi\,\frac{1}{\sqrt{\beta}}\tan^{-1}(\sqrt{\beta}h)
    = \frac{\alpha}{4}\int\!d\psi\,\frac{1}{\sqrt{|\beta|}}\ln\left(\frac{1 + \sqrt{|\beta|}h}{1 - \sqrt{|\beta|}h}\right) ,
 \label{f7c}
 \end{align}
and the integral receives large contributions when $1+\beta h^2\sim 0$, that is, from the ``necks'' of 
the preceding section.  Again by (\ref{e9}), in regions of small $\tilde P$,
\begin{align}
\frac{\alpha}{2}\int\!d\psi\,\frac{1}{\sqrt{\beta}}\tan^{-1}(\sqrt{\beta}h) \approx
   \frac{\alpha}{2}\int\!d\psi\,h\ln\left(\frac{2ff'}{\kappa^2|{\tilde P}|}\right) .
\label{f8}
\end{align}
Classically, we saw in section \ref{sew} that ${\tilde P}\sim\psi^3$ near a neck, while $f'\sim\psi$, so the 
integrand goes as $h\ln(a/h\psi)$ for some parameter $a$.  The integral is maximum when integrated 
over a region of length $\ell_1 \sim a$, and yields
\begin{align}
\ln A\bigl|_{\Delta\ll1}\  \lesssim \  \frac{\alpha\ell_1}{2}  .
\label{f7cc}
\end{align}
(The dependence on the parameter $a$ imposes no further limits.   By section \ref{sew},  
$$a \sim (f^2 D^3K)^{-1/2}\bigl|_{\psi=0}$$
where $K$ is a component of the extrinsic curvature, $D$ is the invariant derivative (\ref{d3}), 
and $f(0)$ is fixed by (\ref{b3}).   If the next order of the expansion (\ref{b4}) involves only constants 
of order one, then $a\sim f_0$, the characteristic size of the neck.  But $f_0$ itself depends on 
the integration constant $\gamma$, and can range from $0$ to $\sqrt{{3}/{\Lambda}}$.  Moreover, 
there is no strong reason to demand that the expansion (\ref{b4}) have coefficients of order one;  
 if extrinsic curvatures remain small near a neck, $a$ can be large.)
 \end{itemize}

Combining these results, we see that the wave function strongly disfavors spaces with large
regions of high extrinsic curvature (large $\Delta$), while giving roughly equal weight to metrics
with regions of relatively small intrinsic curvature near the start of a ``neck'' (trapped surfaces 
$\Delta = 1$) and regions of very small extrinsic curvature where the sign of the expansion can 
change ($\Delta=0$).   Note that at this order, there is no limit to the proper length $\ell$.
Since the integrand in (\ref{f6}) is positive, this suggests an infrared divergence.  As 
observed in section \ref{WQ}, though, we have not yet accounted for the measure in the inner 
product, so it is premature to draw too firm a conclusion.

We can also ask how this higher order WKB term affects the probability current.  For $\beta h^2>-1$,
$A$ is real, so it merely multiplies the lowest order current (\ref{d8}) by $A^2$.  For $\beta h^2<-1$,
the exponent (\ref{f7c}) becomes complex, but the imaginary part is independent of $h$, and does
not contribute to the current.  The moments (\ref{e10}) thus become
\begin{align}
J_n = -\frac{3\kappa^2}{4}\int\!d^3x\,\sigma |A|^2 {\tilde P}\varphi_n .
\label{fx1}
\end{align}
 
 \subsection{Volume regularization}
 
 We can now repeat the argument using volume regularization.  Eqn.\ (\ref{f5}) becomes
 \begin{align}
\frac{1}{A}\frac{\delta A}{\delta h} = \frac{1}{2\ell}\,\frac{1}{1+\beta h^2} ,
\label{f9}
\end{align}
and it is tempting to guess a solution
\begin{align}
A=\exp\left\{ \frac{1}{2\ell}\int\!d\psi\,\frac{1}{\sqrt{\beta}}\tan^{-1}(\sqrt{\beta}h)\right\}  .
\label{f10}
\end{align}
This doesn't quite work, though: because of the metric dependence of $\ell$,  
 \begin{align}
\frac{1}{A}\frac{\delta A}{\delta h} = \frac{1}{2\ell}\,\frac{1}{1+\beta h^2} 
    - \frac{1}{2\ell^2}\int\!d\psi\,\frac{1}{\sqrt{\beta}}\tan^{-1}(\sqrt{\beta}h) .
\label{f11}
\end{align}
In fact, (\ref{f9}) is not integrable:  for any putative solution,
$$\frac{\delta^2A}{\delta h(\psi)\delta h(\psi')} \ne \frac{\delta^2A}{\delta h(\psi')\delta h(\psi)}  .$$
The amplitude (\ref{f10}) is, however, a good \emph{approximate} solution, since it follows from
the preceding section that the extra term in (\ref{f11}) is of order $1/\ell$.   It might be possible to view 
this extra piece as a counterterm in the WKB equation (\ref{f1})---it amounts to adding a term proportional
to $\frac{\delta S}{\delta h}$ to the divergent $\frac{\delta^2 S}{\delta h^2}$---%
but it would be a rather peculiar one.

If we ignore this issue, the conclusions from volume regularization are almost identical
to those from heat kernel regularization.  The only difference is that the prefactor $\alpha$ 
now becomes $1/\ell$, providing a natural infrared cutoff.

\section{Conclusions and next steps}

The space of locally spherically symmetric three-geometries is surprisingly rich.
It includes configurations that exhibit spacetime foam: spaces with multiple layers with 
different geometries, joined by connected sums.  The constraints determine the extrinsic curvature 
only up to a sign, and a typical multilayered configuration includes both expanding and contracting 
regions.  While the these spacetimes are certainly not realistic models of our Universe, they 
are qualitatively very similar to the more general foamy geometries discussed in \cite{Carlip1}.

This structure persists in the quantum theory.  By using the ```dust time'' of Brown and Kucha{\v r}, in
which physical time is proper time along a congruence of timelike geodesics, we can reduce the
Wheeler-DeWitt equation to a Schr{\"o}dinger-type equation, and search for stationary states.  These
states have a sign ambiguity in their phase that closely mimics the classical ambiguity in the sign
of the extrinsic curvature.  When evaluated on a multilayer metric, this sign $\sigma$ can change 
at precisely the locations that the sign of the extrinsic curvature can change classically.  

As we know from quantum mechanics, some care must be taken in interpreting such states.
A stationary state may imply time independence, but it may instead signify a steady state
flow of probability, as in a scattering state.  Indeed, here one can construct a state in which the sign 
$\sigma$ in (\ref{e5}) is the same for every layer.  The probability current is then large, proportional
to the de Sitter expansion $\sqrt{{\Lambda}/{3}}$.  But for a typical state, the current (\ref{fx1}),
evaluated at a typical multilayered geometry, will be very small:  the foamy structure will tend
to reproduce itself in a nearly steady state.  This lends support to the proposal of \cite{Carlip1} that 
a cosmological constant might be ``hidden'' in spacetime foam.  

How common are multilayered foamy configurations?  The second order WKB approximation
has regularization ambiguities, but standard choices such as heat kernel regularization indicate
that they are at least fairly probable.  Indeed, geometries with large regions of high extrinsic curvature
are strongly disfavored, while necks that occur in connected sums of layers are favored.
 Note, though, that the quantum version of the inequality (\ref{a7x}) depends, via (\ref{e2a}), on 
 the state, so for fixed $\gamma$ the prevalence of foamy configurations is state dependent.

There are, of course, further questions that must be answered before we can be confident in
these conclusions.  Perhaps the most important is the problem of the measure for the inner 
product on midisuperspace.  I have been using the ``naive Schr{\"o}dinger interpretation'' for 
$|\Psi|^2$ \cite{Kuchar}, which implicitly assumes that the correct inner product is just a functional 
integral over $h$ and $f$.  But this functional integral must be gauge fixed to account for the
invariances generated by the constraints, and the resulting Faddeev-Popov determinants may 
affect probabilities \cite{Woodard}.   Partial results in this direction are described in the  Appendix.  
Crucially, the Faddeev-Popov determinant is independent of the sign of the extrinsic curvature, 
and will not change the cancellation between expanding and contracting regions. 
 
It would also be useful to move beyond stationary states and look at the behavior of wave packets.
Our choice of time makes this a bit difficult: our ``cloud of clocks'' back-reacts on the 
geometry, and for this to be unimportant we must restrict the energy to be small (on the scale set by
$\Lambda$).  This means wave packets cannot be too sharply peaked.  Ideally, one would
avoid this by choosing a purely gravitational ``internal time,'' but such a parameter seems very
hard to find in a spacetime containing both expanding and contracting regions.  A more careful study 
of the role of the integration constant $\gamma$ would also be valuable.  In particular, it is not 
currently clear whether different choices of $\gamma$ form superselection sectors.

A number of straightforward extensions of this work should be possible.  One could apply the same
techniques to a spatial topology $S^3$, by again starting with a manifold $[0,1]\times S^2$
but now shrinking each boundary to a point.  Beyond that, though, Morrow-Jones and Witt
have shown that local spherical symmetry allows a huge variety of topologies \cite{Witt1},
although a detailed analysis would require much more complicated boundary conditions.
An extension to $\Lambda<0$ should also be easy; most of the computations will be
unchanged, though the qualitative results are likely to be quite different. 

It would also be interesting to look at the ``reduced phase space'' quantization of this model,
in which the reparametrization invariance of $\psi$ is gauge fixed and the momentum constraint
is eliminated classically.  Whether such a quantization is equivalent to the Dirac quantization method used
here is an open question, involving subtleties in operator ordering and the choice of inner 
product  \cite{Kunstatter,Epp}, but it is  certainly worth exploring.  Unfortunately, it is also a bit   
harder than it might appear.  Following \cite{Witt1}, we can \emph{nearly} fix the reparametrization invariance 
by redefining $\psi$ to make $h$ constant on the initial slice.  But the integral $\int\!hd\psi$ is invariant, 
and leaves us with an extra variable (which is also classically time dependent).  The resulting Hamiltonian 
constraint then contains a mixture of ordinary and functional derivatives, and becomes technically quite 
complicated.

Finally, there may be another somewhat orthogonal approach to these questions. 
One way to view spacetime foam in a universe
with large $\Lambda$ is to consider the nucleation of contracting bubbles in expanding regions,
and expanding bubbles in contracting regions.  It is then natural to ask whether
there are instantons that mediate such processes.  In the context of our locally spherically 
symmetric midisuperspace, the question is whether there are Euclidean solutions joining
a space with $n$ layers to one with $n+1$ layers.  Some very preliminary calculations
show no obstruction to such solutions, but much more work is needed.
  
\begin{flushleft}
\large\bf Acknowledgments
\end{flushleft}

 This work was supported in part by Department of Energy grant DE-FG02-91ER40674.
 I would also like to thank the Quantum Gravity Unit of the Okinawa Institute of Science and 
 Technology (OIST), where part of the work was completed, for their hospitality.
 
 \appendix
 \section{A note on the inner product}
 
 As noted in the conclusion, a crucial remaining problem is to find the correct inner product
 on midisuperspace.  Our wave functions are functions of $f$ and $h$, and the naive inner 
 product is simply a functional integral $\displaystyle\int[df][dh]$.  This is not quite right,
 though, for two reasons.
 
 First, our symmetry-reduced midisuperspace still has a residual diffeomorphism invariance,
 reparametrizations of $(t,\psi)$.  The $t$ reparametrizations are fixed by the dust time 
 gauge $t=T$ of section \ref{time}, but the $\psi$ reparametrizations remain, and the integral
 will count identical configurations an infinite number of times.  Second, these gauge fixing of 
 the residual diffeomorphisms will induce a Faddeev-Popov determinant, which must also be
 taken into account.
 
Locally, the $\psi$ reparametrizations can be fixed by setting $\partial_\psi h=0$, as was
done in \cite{Witt1}.  Note that the total length $\int\! d\psi\, h$ is diffeomorphism invariant,
so the constant value of $h$ is fixed.  Whether this choice can be made globally over the
whole midisuperspace---that is, whether there are Gribov problems \cite{Gribov,Schoen}---is  
a more difficult question.

The Faddeev-Popov determinants are best understood as Jacobians in the functional integral, 
as described in \cite{Polchinski,CMNP}.  (For an early version, see \cite{DeWitt2}.)  Let $\phi$ 
be a generic field, and define an inner product $\langle\delta\phi,\delta\phi\rangle$ on the
tangent space to the space of fields.  The standard normalization for the path integral is
\begin{align}
\int [d\phi]e^{-\langle\delta\phi,\delta\phi\rangle} = 1   .
\label{A1}
\end{align}
But now suppose $\phi$ has a gauge symmetry $\phi\rightarrow{}^\eta\phi$, labeled by some 
parameter $\eta$.  We can change variables in the functional integral to $({\bar\phi},\eta)$, 
where $\bar\phi$ is a gauge-fixed field.  Geometrically, $\eta$ parametrizes gauge orbits,
while $\bar\phi$ parametrizes a cross section that intersects each orbit once.  (Whether such
a cross section exists or not is the Gribov problem.)  Then
\begin{align}
\int [d\phi]e^{-\langle\delta\phi,\delta\phi\rangle}  
   = \int [d{\bar\phi}][d\eta] Je^{-\langle\delta{\bar\phi},\delta{\bar\phi}\rangle}%
   e^{-\langle\delta_\eta\phi,\delta_\eta\phi\rangle}   
   = \int [d\eta] J e^{-\langle\delta_\eta\phi,\delta_\eta\phi\rangle} = 1 ,
\label{A2}
\end{align}
where $J$ is a Jacobian coming from the change of variables.  This Jacobian is the
Faddeev-Popov determinant.

For us, the fields are the metric $q_{ij}$ and the ``dust'' field $T$, with natural inner products
\begin{align}
\langle\delta g,\delta g\rangle &= \int\!d^3x\,\sqrt{q}\,q^{ij}q^{k\ell}\delta q_{ik}\delta q_{j\ell}
   = 4\pi \int d\psi\,hf^2\left[ 4\left(\frac{\delta h}{h}\right)^2 + 8\left(\frac{\delta f}{f}\right)^2\right] \nonumber\\[1ex]
\langle\delta T,\delta T\rangle &= \int\!dx^3\,\sqrt{q}\,(\delta T)^2
  = 4\pi \int d\psi\,hf^2(\delta T)^2   .
\label{A3}
\end{align} 
The transformations generated by the constraints (\ref{a0a})--(\ref{a0b}) involve 
two parameters ($\xi,\xi^\perp$), a $\psi$ reparametrization and a ``surface deformation'' (basically a $t$
reparametrization).  It is straightforward to check that
\begin{align}
\frac{\delta_\eta h}{h} = \xi' + \frac{h'}{h}\xi + K^\psi{}_\psi\xi^\perp ,\quad
\frac{\delta_\eta f}{f} =  \frac{f'}{f}\xi + K^\theta{}_\theta\xi^\perp ,\quad
\delta_\eta T = \xi^\perp ,  
\label{A4}
\end{align} 
where $K^\psi{}_\psi$ and $K^\theta{}_\theta$ are the extrinsic curvatures (\ref{a3a}).  Thus the
inner product $\langle\delta_\eta\phi,\delta_\eta\phi\rangle$ becomes 
\begin{align}
\langle\delta_\eta\phi,\delta_\eta\phi\rangle 
   &=  4\pi\int d\psi\,hf^2 \left[ 4\left(\xi' + \frac{h'}{h}\xi + K^\psi{}_\psi\xi^\perp\right)^2
       + 8\left(\frac{f'}{f}\xi + K^\theta{}_\theta\xi^\perp\right)^2 + (\xi^\perp)^2\right]   . 
\label{A5}
\end{align}       
Let us define
$${\tilde K}^2 = K^i{}_jK^j{}_i = K^\psi{}_\psi{}^2 + 2 K^\theta{}_\theta{}^2$$
Then the integrand in (\ref{A5}) has the form
$$
(1+4{\tilde K}^2)(\xi^\perp)^2 
   + \left[ 8\left(\xi' + \frac{h'}{h}\xi \right)K^\psi{}_\psi + 16 \frac{f'}{f}\xi K^\theta{}_\theta\right]\xi^\perp + \dots
$$
where the remaining terms are independent of $\xi^\perp$.  We can now complete the square, setting
$$
{\bar\xi}^\perp = \xi^\perp + 4 (1+4{\tilde K}^2)^{-1}
\left[\left(\xi' + \frac{h'}{h}\xi \right)K^\psi{}_\psi + 2 \frac{f'}{f}\xi K^\theta{}_\theta\right]  .
$$
After some integrations by parts, we obtain an expression of the form
\begin{align}
\langle\delta_\eta\phi,\delta_\eta\phi\rangle 
   =  4\pi\int d\psi\,hf^2 \left[ ({\bar\xi}^\perp)^2 + B\xi(D^2+V)\xi\right]  ,
\label{A6}
\end{align}
where $D$ is the derivative (\ref{d3}).  
The coefficients $B$ and $V$ are complicated functions of $f$ and $h$, but they are independent of 
${\bar\xi}^\perp$.  Crucially, they are also quadratic in the extrinsic curvature, and hence invariant 
under $K^i{}_j\rightarrow -K^i{}_j$.  The Jacobian (\ref{A2}) thus takes the form 
\begin{align}
J = \det{}^{1/2}|B(D^2+V)| ,
\label{A7}
\end{align}
and is independent of the sign of the extrinsic curvature.  Hence the inner product does not
break the time-reversal symmetry that produces cancellations between regions with different
signs $\sigma$.

It may be possible to obtain further information about $J$ through zeta function methods.  In 
particular, the scaling behavior could be important in understanding the possible infrared divergences
discussed in section \ref{heat}  For now, though, this lies beyond the scope of this paper.


\small

 \begin{thebibliography}{99}
\bibitem{Wheeler} J.~A.\ Wheeler, ``Geons,'' Phys.\ Rev.\ 97 (1955) 511.
\bibitem{Carlip1} S.\ Carlip, ``Hiding the Cosmological Constant,'' Phys.\ Rev.\ Lett.\ 123 (2019) 
131302, arXiv:1809.08277.
\bibitem{Chrusciel} P.~T.\ Chrusciel, J.\ Isenberg, and D.\ Pollack, ``Gluing initial data sets 
for general relativity,'' Phys.\ Rev.\ Lett.\ 93 (2004) 081101, arXiv:gr-qc/0409047.
\bibitem{Chruscielb} P.~T.\ Chrusciel, J.\ Isenberg, and D.\ Pollack, ``Initial data engineering,''
Commun.\ Math.\ Phys.\ 257 (2005) 29, arXiv:gr-qc/0403066.
\bibitem{Witt1} J.\ Morrow-Jones and D.~M.\ Witt, ``Inflationary initial data for generic spatial topology,''
 Phys.\ Rev.\ D 48 (1993) 2516.
 \bibitem{Witt2} K.\ Schleich and D.~M.\ Witt, ``Designer de Sitter Spacetimes,'' Can.\ J.\ Phys.\ 86 (2008) 591.
 arXiv:0807.4559.
 \bibitem{Witt3} K.\ Schleich and D.~M.\ Witt, ``What does Birkhoff's theorem really tell us?'' 
 arXiv:0910.5194.
 \bibitem{Brown} J.~D.\ Brown and K.~V.\ Kucha{\v r}, ``Dust as a standard of space and time 
in canonical quantum gravity,'' Phys.\ Rev.\ D51 (1995) 5600, arXiv:gr-qc/9409001.
\bibitem{Marolf} J.~D.\ Brown and D.\ Marolf, ``On relativistic material reference systems,''
Phys.\ Rev.\ D53 (1996) 1835, arXiv:gr-qc/9509026.
 \bibitem{Husain} V.\ Husain and T.\ Pawlowski, ``Time and a physical Hamiltonian for quantum gravity,''
 Phys.\ Rev.\ Lett.\ 108 (2012) 141301, arXiv:1108.1145.
 \bibitem{Carlip3} S.\ Carlip, \emph{General Relativity} (Oxford University Press, 2019).
 \bibitem{Lake} K.\ Lake and R.~C.\ Roeder, ``Effects of a Nonvanishing Cosmological Constant on the 
 Spherically Symmetric Vacuum Manifold,''  Phys.\ Rev.\ D 15 (1977) 3513.
 \bibitem{SW} K.\ Schleich and D.~M.\ Witt, ``A simple proof of Birkhoff's theorem for cosmological constant,'' 
J.\ Math.\ Phys.\ 51 (2010) 112502, arXiv:0908.4110.
\bibitem{Burkhartb} M.\ Burkhart and D.\ Pollack, ``Causal geodesic incompleteness of spacetimes
arising from IMP gluing,'' arXiv:1907.00295.
\bibitem{Kuchar} K.~V.\ Kucha{\v r}, ``Time and Interpretations of Quantum Gravity,'' 
Int.\ J.\ Mod.\ Phys.\ D20 (2011) 3.
\bibitem{Isham} C.~J.\ Isham, ``Canonical quantum gravity and the problem of time,'' NATO Sci.\
Ser.\ C 409 (1993) 157, arXiv:gr-qc/9210011.
\bibitem{Misner} C.~W.\ Misner, ``Quantum cosmology 1,'' Phys.\ Rev.\ 186 (1969) 1319.
\bibitem{expand} N.~{\'O} Murchadha, C.\ Soo, and H.-L.\ Yu, ``Intrinsic time gravity and the 
Lichnerowicz-York equation,'' Class.\ Quant.\ Grav.\ 30 (2013) 095016, arXiv:1208.2525.
\bibitem{York} Y.\ Choquet-Bruhat and J.\ York, ``The Cauchy problem,'' in  \emph{General 
Relativity and Gravitation I}, edited by A.\ Held (New York: Plenum, 1980).
\bibitem{MCF} M.\ Kleban and L.\  Senatore, ``Inhomogeneous Anisotropic Cosmology,'' 
JCAP 10 (2016) 022, arXiv:1602.03520.
\bibitem{costime} L.\ Andersson, G.~J.\ Galloway, and R.\ Howard, ``The Cosmological time function,''
Class.\ Quant.\ Grav.\ 15 (1998) 309, arXiv:gr-qc/9709084.
\bibitem{Torre} K.~V.\ Kuchar and C.~G.\ Torre, ``Gaussian reference fluid and interpretation of 
quantum geometrodynamics,''  Phys.\ Rev.\ D 43 (1991) 419.
\bibitem{unimodular} W.~G.\ Unruh, ``A Unimodular Theory of Canonical Quantum Gravity,''
Phys.\ Rev.\ D 40 (1989) 1048.
\bibitem{Woodard} R.~P.\ Woodard, ``Enforcing the Wheeler-de Witt Constraint the Easy Way,''
 Class.\ Quant.\ Grav.\ 10 (1993) 483.
 \bibitem{WdW} B.~S.\ DeWitt, ``Quantum Theory of Gravity 1. The Canonical Theory,''
Phys.\ Rev.\ 160 (1967) 1113.
\bibitem{Marolfb} D.\ Marolf, ``Group averaging and refined algebraic quantization: Where are we now?''
in \emph{Proc.\ of the 9th Marcel Grossmann Meeting}, edited by V.~G.\ Gurzadian, R.~T.\ Jantzen,
and R.\ Ruffini (Singapore: Wold Scientific, 2002), arXiv:gr-qc/0011112.
\bibitem{Feng} J.~C.\ Feng, ``Volume average regularization for the Wheeler-DeWitt equation,''
Phys.\ Rev.\ D 98 (2018) 026024, arXiv:1802.08576.
\bibitem{Horiguchi} T.\ Horiguchi(, K.\ Maeda(, and M.\ Sakamoto, ``Analysis of the Wheeler-DeWitt 
equation beyond Planck scale and dimensional reduction,'' Phys.\ Lett.\ B 344 (1995) 105,
arXiv:hep-th/9409152.
\bibitem{Kunstatter} G.\ Kunstatter, ``Dirac versus reduced quantization: A Geometrical approach,''
Class.\ Quant.\ Grav.\ 9 (1992) 1469.
\bibitem{Epp} R.~J.\ Epp, ``Curved space quantization: Toward a resolution of the Dirac versus reduced quantization 
question,'' Phys.\ Rev.\ D 50 (1994) 6578.
\bibitem{Gribov}  V.~N.\ Gribov, ``Quantization of Nonabelian Gauge Theories,''  Nucl.\ Phys.\ B 139 
(1978) 1.
\bibitem{Schoen} M.\ Schon and p.\  Hajicek, ``Topology of Quadratic Superhamiltonians,''
Class.\ Quant.\ Grav.\ 7 (1990) 861.
\bibitem{Polchinski} J.\ Polchinski, ``Evaluation of the One Loop String Path Integral,''  Commun.\
Math.\ Phys.\ 104 (1986) 37.
\bibitem{CMNP} A.~G.\ Cohen, G.~W.\ Moore, P.~C.\ Nelson, and J.\ Polchinski, ``An Off-Shell Propagator 
for String Theory,''  Nucl.\ Phys.\ B 267 (1986) 143.
\bibitem{DeWitt2} B.~S.\ DeWitt, ``Quantum Theory of Gravity 2. The Manifestly Covariant Theory,''
Phys.\ Rev.\ 160 (1967) 1195.
 
 
 \end{thebibliography}
\end{document}